%% file: paper.tex
\documentclass[letter,12pt]{article}

\usepackage{graphicx}
\usepackage{scicite}
\usepackage{times}
\usepackage{natbib}
\usepackage{fancyhdr}
\usepackage{multirow}

\newenvironment{sciabstract}{%
\begin{quote} \bf}
{\end{quote}}

\title{
Classes of complex networks defined \\
by role-to-role connectivity profiles
}
\author{Roger Guimer\`a, Marta Sales-Pardo, and Lu\'{\i}s A. N.
Amaral\\
\normalsize{Department of Chemical and Biological Engineering and }\vspace{-.3cm}\\
\normalsize{Northwestern Institute on Complex Systems (NICO)}\vspace{-.3cm}\\
\normalsize{Northwestern University, Evanston, IL 60208, USA}
}

\date{}

\renewcommand{\baselinestretch}{1.5}

\begin{document}

\maketitle
%


\begin{sciabstract}
\input{abstract}
\end{sciabstract}



The structure of complex networks~\cite{newman03,amaral04} is
typically characterized in terms of global properties, such as the
average shortest path length between nodes~\cite{watts98}, the
clustering coefficient~\cite{watts98}, the
assortativity~\cite{newman02} and other measures of degree-degree
correlations~\cite{pastor-satorras01b,colizza06}, and, especially, the
degree distribution~\cite{barabasi99,amaral00}. %
However, these global quantities are truly informative only when one
of two strict conditions is fulfilled: (i) the network lacks a modular
structure~\cite{girvan02,newman04b,guimera05a,guimera05,guimera05b,danon05},
or (ii) the network has a modular structure but (ii.a) all modules
were formed according to the same mechanisms, and therefore have
similar properties, and (ii.b) the interface between modules is
statistically similar to the bulk of the modules, except for the
density of links. If neither of these two conditions is fulfilled,
then any theory proposed to explain, for example, a scale-free degree
distribution needs to take into account the modular structure of the
network.


To our knowledge, no real-world network has been shown to fulfill
either of the two conditions above; this implies that global
properties may sometimes fail to provide insight into the mechanisms
responsible for the formation or growth of these networks. Alternative
approaches that take into consideration the modular structure of
real-world complex networks are therefore necessary.
One such approach is to group nodes into a small number of roles,
according to their pattern of intra- and inter-module
connections~\cite{guimera05a,guimera05,guimera05b}. Recently, we
demonstrated that the role of a node conveys significant information
about the importance of the node, and about the evolutionary pressures
acting on it~\cite{guimera05a,guimera05b}.
%
%
Here, we demonstrate that modular networks can be classified into
distinct functional classes according to the patterns of role-to-role
connections, and that the definition of link types can help us
understand the function and properties of a particular class of
networks.


\section*{Modularity of complex networks}


We analyze four different types of real-world networks---metabolic
networks~\cite{guimera05a,jeong00,wagner01}, protein
interactomes~\cite{uetz00,jeong01,maslov02,li04}, global and regional
air transportation networks~\cite{guimera05b,barrat04,li04a}, and the
Internet at the autonomous system (AS)
level~\cite{pastor-satorras01b,vazquez02} (Table~\ref{t-datasets} and
Supplementary discussion). To determine and quantify the modular
structure of these networks, we use simulated
annealing~\cite{kirkpatrick83} to find the optimal partition of the
network into modules~\cite{guimera05a,guimera05,guimera04}
(Methods). We then assess the significance of the modular structure of
each network by comparing it to a randomization of the same
network~\cite{guimera04}. We find that all networks studied have a
significant modular structure (Table~\ref{t-datasets}). Modules
correspond to functional units in biological
networks~\cite{guimera05a,li04} and to geo-political units in air
transportation networks~\cite{guimera05b} and, probably, in the
Internet~\cite{eriksen03}.

To assess whether global average properties are appropriate to
describe the structure of these networks, we compare global average
properties of the networks to the corresponding module-specific
averages; specifically, we focus on the degree, the clustering
coefficient, and the normalized clustering coefficient. We find that
the average degree of the network is not representative of
individual-module average degrees for air transportation networks
(Table~ \ref{t-globprop}). Most importantly, the global clustering
coefficient is not representative of individual-module clustering
coefficients for any network (except, maybe, for one out of 18
metabolic networks).

\section*{Role-based description of complex networks}


As an alternative to the average description approach, we determine
the role of each node according to two properties
\cite{guimera05a,guimera05} (Methods): the relative within-module
degree $z$, which quantifies how well connected a node is to other
nodes in their module, and the participation coefficient $P$, which
quantifies to what extent the node connects to different modules. We
classify as {\it non-hubs} those nodes that have low within-module
degree ($z<2.5$). Depending on the fraction of connections they have
to other modules, non-hubs are further subdivided
into~\cite{guimera05a,guimera05}: (R1) {\it ultra-peripheral nodes},
that is, nodes with all their links within their own module; (R2) {\it
peripheral nodes}, that is, nodes with most links within their module;
(R3) {\it satellite connectors}, that is, nodes with a high fraction
of their links to other modules; and (R4) {\it kinless nodes}, that
is, nodes with links homogeneously distributed among all modules. We
classify as {\it hubs} those nodes that have high within-module degree
($z\ge 2.5$). Similar to non-hubs, hubs are divided according to their
participation coefficient into: (R5) {\it provincial hubs}, that is,
hubs with the vast majority of links within their module; (R6) {\it
connector hubs}, that is, hubs with many links to most of the other
modules; and (R7) {\it global hubs}, that is, hubs with links
homogeneously distributed among all modules.

Although the full rationale for this particular definition of the
roles has been given elsewhere~\cite{guimera05}, it is important to
highlight a few properties of our classification scheme. Nodes in real
and model networks, especially non-hubs, do not fill uniformly the
$zP$-plane; our role classification scheme arises from the fact that
nodes tend to congregate into a small number of densely populated
regions of this space, with boundaries between these regions having
low density of nodes. Additionally, especially for hubs, boundaries
coincide with well defined connectivity patterns; for example, nodes
at the boundary between connector hubs (R6) and global hubs (R7) would
have approximately half of their links in one module, and the other
half perfectly spread in other modules. Importantly, other definitions
of the roles do not alter the results we report below (see
Supplementary Information).


We investigate how our definition of roles relates to global network
properties, and to what extent global network properties are
representative of nodes with different roles. Since some simple
properties like the degree and the clustering coefficient trivially
depend on a node's role, we focus on degree-degree
correlations~\cite{newman02,pastor-satorras01b,maslov02,park03,maslov04,colizza06}. Specifically,
we address two questions: (i) whether nodes with the same degree but
different roles have the same or different correlations; and (ii) to
what extent the observed degree-degree correlations are a byproduct of
the modular structure of the network.

\begin{figure}
\centerline{
\includegraphics*[width=\textwidth]{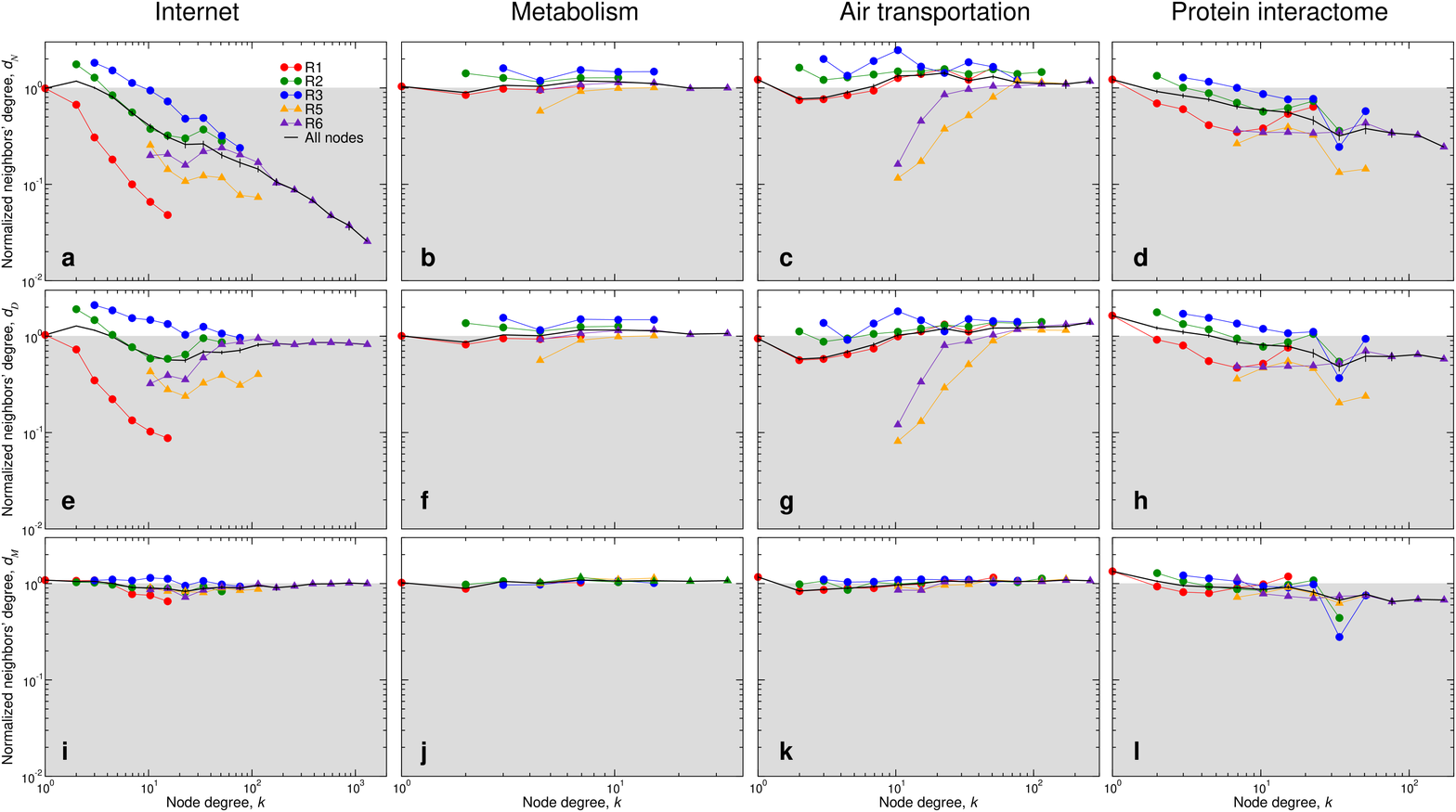}
}
\caption{
Modularity and degree distribution explain most degree-degree
correlations in complex networks.
{\bf a-d}, Degree $d_\mathcal{N}$ of the neighbors of a node
normalized by the average neighbors' degree of all the nodes in the
network;
{\bf e-h}, Degree $d_\mathcal{D}$ of the neighbors of a node
normalized by the average neighbors' degree of the node in the
ensemble of random networks with fixed degree sequence; and
{\bf i-l}, Neighbors' degree $d_\mathcal{M}$ of a node normalized by
the average neighbors' degree of the node in the ensemble of random
networks with fixed degree sequence and modular structure (Methods).
Values of $d$ are averaged over nodes with similar degree to obtain
the function $d(k)$. Error bars represent the standard error of the
average.
Note that a lack of deviations from the ensemble average, that is,
$d(k) = 1$, indicates the absence of correlations. The results in the
middle row show that the degree distribution is responsible for some
of the observed degree-degree correlations, but cannot fully account
for them. The degree distribution and the modular structure of the
network do account for most existing degree-degree correlations in the
Internet, metabolic and air transportation networks.
}
\label{f-knn}
\end{figure}

To answer these questions, we start by considering the Internet at the
AS level (Fig.~\ref{f-knn}). Nodes with degree $k=3$ can be either
ultra-peripheral (R1, if they have all connections in the same
module), peripheral (R2, if they have two connections in one module
and one in another), or satellite connectors (R3, if the three
connections are to different modules). A separate analysis for each
role reveals that the average degree $k^{\rm nn}(k)$ of the neighbors
of a node~\cite{pastor-satorras01b} with degree $k=3$ strongly depends
on the role of the node. For an instance of the 1998 Internet, for
example, $k^{\rm nn}(k=3)=43 \pm 8$ for ultra-peripheral nodes,
$k^{\rm nn}(k=3)=196 \pm 12$ for peripheral nodes, and $k^{\rm
nn}(k=3)=290 \pm 20$ for satellite connectors. We observe a dependence
of $k_{\rm nn}$ on the nodes' role for all the networks studied here
(Fig.~\ref{f-knn}a-d).

Regarding the second question, initial research
showed~\cite{pastor-satorras01b} that for the Internet at the AS level
$k^{\rm nn}(k)\propto k^{-0.5}$. It was later pointed
out~\cite{maslov04,park03} that {\it any} network with the same degree
distribution as the Internet should display a similar scaling. In
other words, the degree distribution of the network is responsible for
most of the observed correlations. However, the degree distribution
alone does not account for all the observed
correlations~\cite{maslov04} (Fig.~\ref{f-knn}e). In contrast, the
modular structure of the network does account for most of the
remaining degree-degree correlations observed in the topology of the
Internet (Fig.~\ref{f-knn}i). Similarly, the modular structure
accounts for the degree-degree correlations in metabolic networks and
the air transportation network, and for most of the correlations in
protein interaction networks (Fig.~\ref{f-knn}i-l).

\section*{Role-to-role connectivity profiles}

The findings we reported so far suggest that, once the degree
distribution and the modular structure are fixed, real networks have
no additional internal structure. This, however, contradicts our
intuition that {\it networks with different growth mechanisms and
functional needs should have distinct connection patterns between
nodes playing different roles}. To investigate this possibility, we
systematically analyze how nodes connect to one another depending on
their roles.

For each network, we calculate the number $r_{ij}$ of links between
nodes belonging to roles $i$ and $j$, and compare this number to the
number of such links in a properly randomized network (Methods). As in
previous work~\cite{maslov02,milo02,maslov04,milo04}, we use the
$z$-score to obtain a profile $\vec{a}$ of over- and
under-representation of link types (Fig.~\ref{f-r2r}), which enables
us to compare different networks. We quantify the overall similarity
between two profiles $\vec{a}$ and $\vec{b}$ by the scalar product
between these profiles (Methods). In Fig.~\ref{f-r2r}, we show that
networks of the same type have highly correlated profiles, while
networks of different types have weaker correlations and, at times,
even strong anti-correlations (Fig.~\ref{f-r2r}c).

The networks considered fall into two main classes, one comprising
metabolic and air transportation networks, and another comprising
protein interactomes and the Internet. The main difference between the
two groups is the pattern of links between: (i) ultra-peripheral nodes
(links of type R1-R1), and (ii) connector hubs and other hubs (links
of types R5-R6 and R6-R6). These link types are over-represented for
networks in the first class (except links of type R6-R6 in metabolic
networks), and under-represented for networks in the second class.

\begin{figure}
\centerline{
\includegraphics*[width=\textwidth]{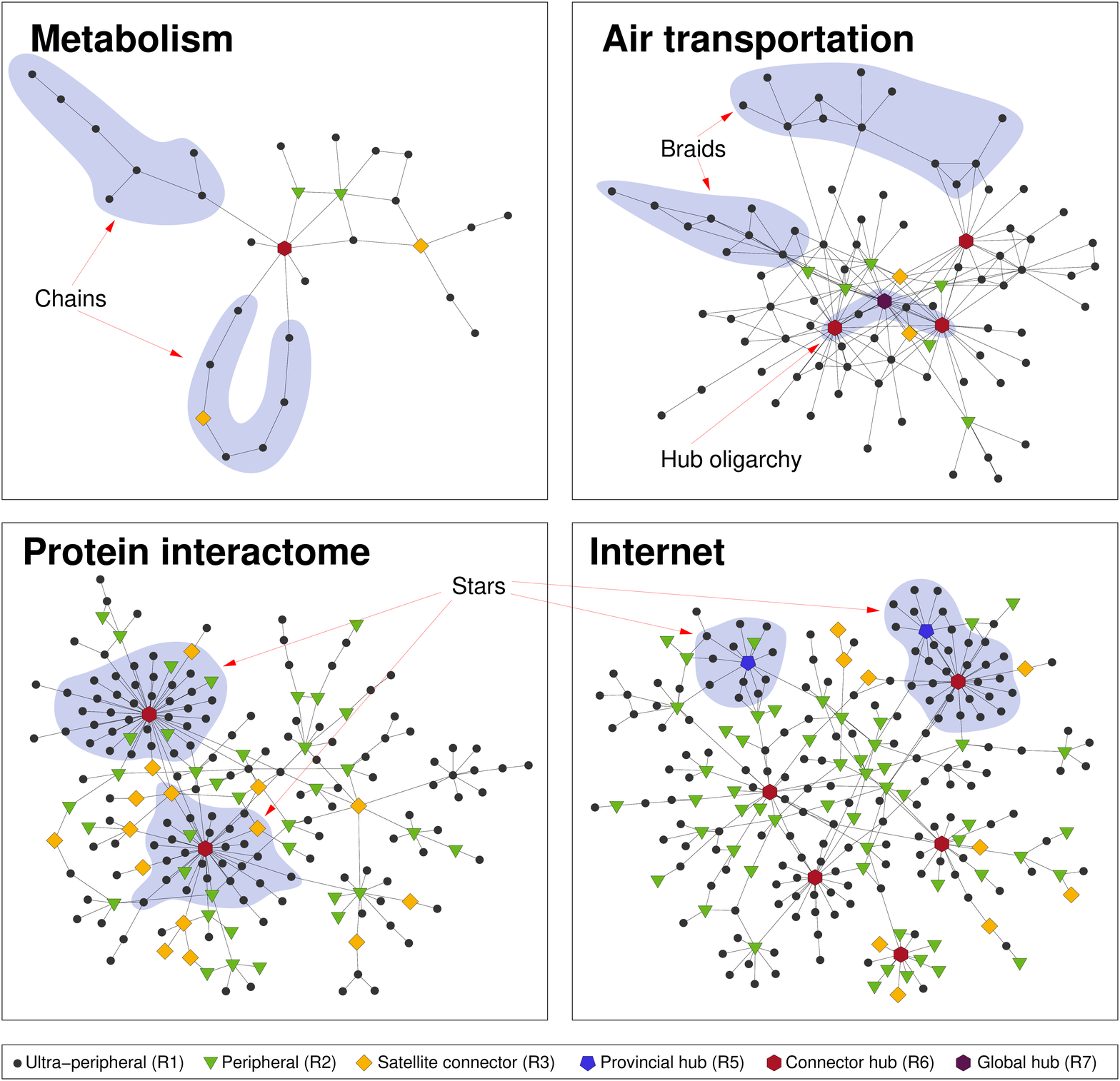}
}
\caption{
Modules and role-to-role connectivity signatures in different network
types. Each panel represents a single module (that is, all the nodes
depicted belong to a single module) in: the metabolic network of {\it
A. thaliana}, the Asia and Middle East air transportation network, the
protein interactome of {\it C. elegans}, and the Internet in 1998.
Different symbols indicate different node roles (see Supplementary
Discussion for the names of the nodes). External links to other
modules are not depicted, although it is possible to infer where they
are from the role of the nodes. Shaded regions highlight important
structural features.
}
\label{f-modules}
\end{figure}

We denote the first class as the {\it stringy-periphery} class
(Fig.~\ref{f-modules}a, b). In networks of this class,
ultra-peripheral nodes are more connected to one another than one
would expect from chance, which results in long ``chains'' of
ultra-peripheral nodes. In metabolic networks, these chains correspond
to loop-less pathways that, for example, degrade a complex metabolite
into simpler molecules. In the air transportation network, due to the
higher overall connectivity of the network, chains contain short loops
and resemble ``braids.'' Stringy-periphery networks also have a core
of hubs, which we call the {\it hub oligarchy}, that are directly
reachable from one another (links of type R5-R6 in metabolic and air
transportation networks, and R6-R6 in air transportation
networks). Moreover, connector hubs are less connected to
ultra-peripheral nodes (R1) than expected by chance alone.


We denote the second class as the {\it multi-star} class
(Fig.~\ref{f-modules}c, d). The multi-star class comprises the protein
interactomes and the Internet, and has the opposite signature to the
stringy-periphery class. Links of type R1-R1 (between ultra-peripheral
nodes) are under-represented, whereas links of type R1-R5 (between
ultra-peripheral nodes and provincial hubs) are, over-represented,
giving rise to modules with indirectly-connected ``star-like''
structures. Similarly, connector hubs are less connected to one
another than one would expect, which means that these networks depend
on satellite connectors to bridge connector hubs and modules.

Our findings confirm and clarify previous results in the
literature. For example, the under-repre\-sentation of R6-R6 links in
protein interactomes is consistent with previous results suggesting a
tendency for hubs to ``repel'' each other in these
networks~\cite{maslov02,colizza06}. Similarly, the role-to-role
connectivity profile of the Internet is consistent with the existence
of a hierarchy of types of nodes~\cite{maslov04}. This hierarchy
comprises end users, regional providers, and global providers, which
we hypothesize correspond correspond to roles R1-R2, R5, and R6
respectively. The role-to-role connectivity profiles are consistent
with a scenario in which end users connect mostly to regional
providers, and in which global providers connect with each other
indirectly through satellite connectors (R3), with few connections but
probably large bandwidth.

By considering the modular structure of the networks and the extra
dimension introduced by the participation coefficient, however, our
approach provides novel insights into the relationship between
structure and function in complex networks. For example, by
considering the absolute degree alone nodes with roles R5 and R6 in
protein interactomes are indistinguishable from each other: in {\it
S. cerevisiae}, $\langle k \rangle_{R5}=14.0 \pm 1.7$ and $\langle k
\rangle_{R6}=17.1 \pm 1.9$, whereas the average degree for the whole
network is $\langle k \rangle=2.67 \pm 0.09$. Still, links R5-R5
between provincial hubs, unlike R6-R6 links, are not
under-represented. In general, the different connection patterns of R5
and R6 (or R1 and R2) proteins enables us to hypothesize that they
play distinct biological roles, with R6 proteins likely being much
more important~\cite{han04}.

A closer look at the air transportation network also helps to show
that important structural properties may be left unexplained by
focusing on degree alone, as well as to stress the importance of the
relative within-module degree as opposed to the degree. Johannesburg,
in South Africa, has degree $k=$84, which is 23\% smaller than the
degree of Cincinnati in the U.S., $k=$109. Still, one can fly from
most capitals in the world to Johannesburg but not to
Cincinnati. There are two main reasons for this. First, while
Johannesburg is the most connected city in its region (sub-Saharan
Africa), Cincinnati (North America) is not; this effect is captured by
the within-module relative degree, which is 9.3 for Johannesburg and
4.3 for Cincinnati. Second, Johannesburg has many connections to other
regions, whereas Cincinnati does not; this effect is captured by the
participation coefficient, which is 0.52 for Johannesburg and 0.05 for
Cincinnati. As a result, Johannesburg is a global hub (R6) in our
classification, whereas Cincinnati is a provincial hub (R5). One can
thus understand why R6-R6 connections are over-represented in air
transportation networks (most global hubs are connected to one
another), whereas R5-R5 are not (most provincial hubs are poorly
connected to provincial hubs in other regions). In general, our
approach shows why the behavior of R5 and R6 nodes is so different in
air transportation networks, which cannot be understood from the
degree of the nodes alone.

\section*{Conclusion}

We have shown that global properties that do not take into account the
modular organization of the network may sometimes fail to capture
potentially important structural features; although all networks
(except, maybe, the protein interactomes) show no degree-degree
correlations when compared to the appropriate ensemble of random
networks, they all have clearly distinctive properties in terms of how
nodes with certain roles are connected to each other. Our results thus
call attention to the need to develop new approaches that will enable
us to better understand the structure and evolution of real-world
complex networks.

Additionally, our findings demonstrate that networks with the same
functional needs and growth mechanisms have similar patterns of
connections between nodes with different roles. Attempts to divide
complex networks into ``classes'' or ``families'' have been made
before, for example in terms of the degree
distribution~\cite{amaral00} and in terms of the relative abundance of
certain subgraphs or motifs~\cite{milo02,milo04}. Our work here
complements those attempts, and is the first one to build on the
crucial fact that most real-world networks display a markedly modular
structure.

Although we cannot put forward a theory for the division of the
networks into two classes, we hypothesize that it might be related to
the fact that networks in the stringy-periphery class are
transportation networks, in which strict conservation laws must be
fulfilled. Indeed, for transportation systems it has been shown that,
under quite general conditions, a {\it hub oligarchy} is the the most
efficient organization~\cite{arenas03}. Conversely, both protein
interactomes and the Internet can be seen as signaling networks, which
do not obey conservation laws.


\include{methods}

%
%
\bibliography{ref-database}

\bigskip
\noindent
Correspondence and requests for materials should be addressed to R.~G.

\bigskip
\noindent
{\bf Acknowledgments}~~We thank R.D. Malmgren, E.N. Sawardecker,
S.M.D. Seaver, D.B. Stouffer, and M.J. Stringer for useful comments
and suggestions. R.G. and M.S.-P. thank the Fulbright
Program. L.A.N.A. gratefully acknowledges the support of a NIH/NIGMS
K-25 award, of NSF award SBE 0624318, of the J.S. McDonnell
Foundation, and of the W.~M. Keck Foundation.

\clearpage

\renewcommand{\baselinestretch}{1.0}

\input{table-datasets}

\input{table-module-globprop-likelyhood}

\clearpage


\begin{figure}
\vspace{-1cm}
\centerline{
\includegraphics*[width=\textwidth]{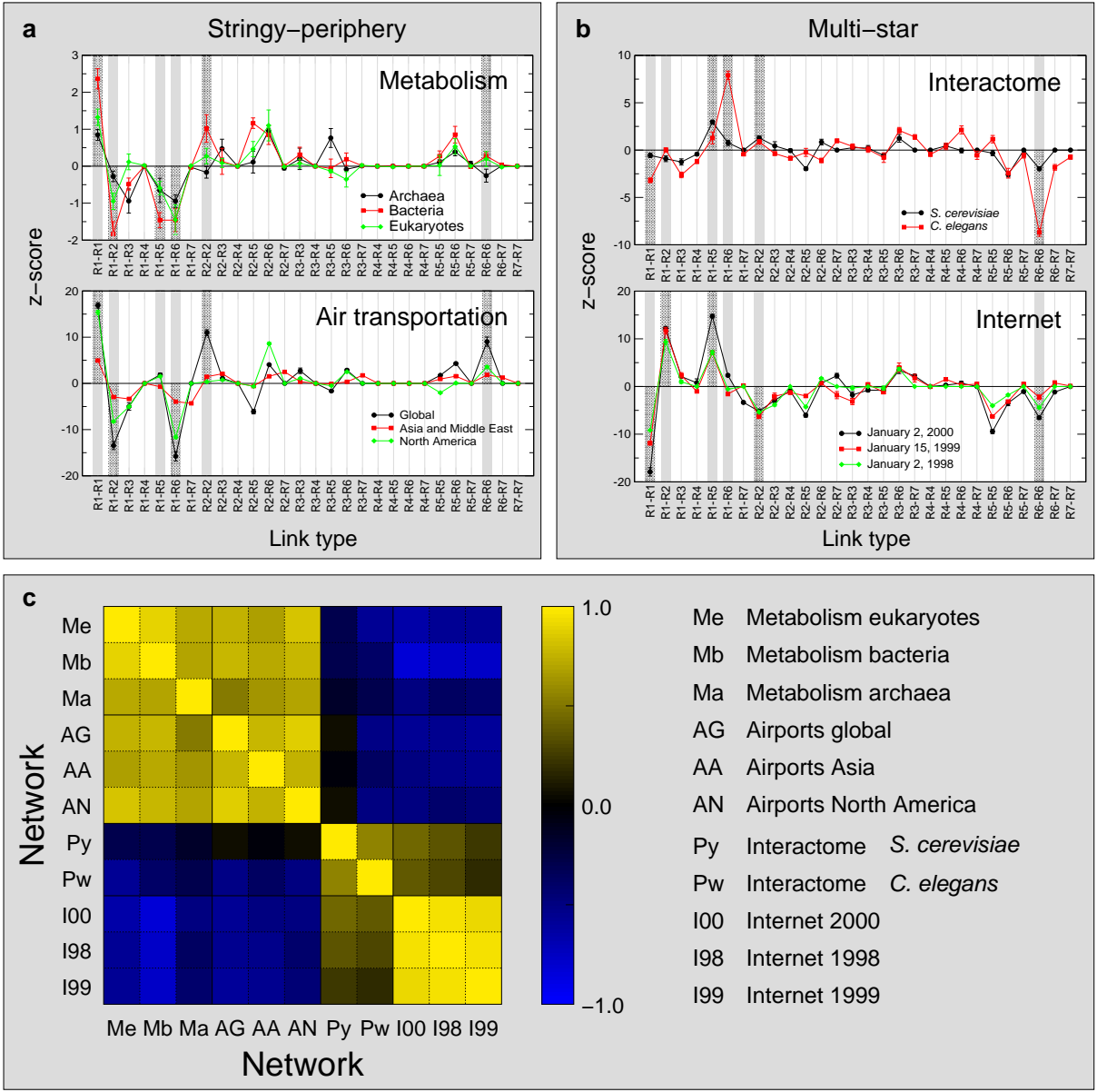}
}
\renewcommand{\baselinestretch}{1.0}
\caption{
Role-to-role connectivity patterns.
We plot the $z$-score for the abundance (Methods) of each link type
for: {\bf a}, {\it stringy-periphery} networks, and {\bf b}, {\it
multi-star} networks (see text). Roles are labeled as follows: (R1)
ultra-peripheral; (R2) peripheral; (R3) satellite connectors; (R4)
kinless nodes; (R5) provincial hubs; (R6) connector hubs; (R7) global
hubs.
{\bf c}, We quantify the similarity between two $z$-score profiles by
means of the correlation coefficient (Methods), with yellow
corresponding to large positive correlation, blue to large
anti-correlation, and black to no correlation.
%
%
Gray columns in {\bf a} indicate those link types that contribute the
most, in absolute value, to the correlation coefficient. These link
types are, therefore, the ones that better characterize the set of all
profiles.
}
\label{f-r2r}
\end{figure}


\end{document}

%% file: abstract.tex
Interactions between units in phyical, biological, technological, and social systems usually give rise to intrincate networks with non-trivial structure, which critically affects the dynamics and properties of the system. The focus of most current research on complex networks is on global network properties. A caveat of this approach is that the relevance of global properties hinges on the premise that networks are homogeneous, whereas most real-world networks have a markedly modular structure. Here, we report that networks with different functions, including the Internet, metabolic, air transportation, and protein interaction networks, have distinct patterns of connections among nodes with different roles, and that, as a consequence, complex networks can be classified into two distinct functional classes based on their link type frequency. Importantly, we demonstrate that the above structural features cannot be captured by means of often studied global properties.

%% file: methods.tex
\section*{Methods}


\subsection*{Module identification}

The modularity $\mathcal{M(P)}$ of a partition $\mathcal{P}$ of a
network into modules is~\cite{newman04b}
\begin{equation}
\mathcal{M(P)}\equiv\sum_{s=1}^{N_M}\left[\frac{l_{s}}{L}-
\left(\frac{d_s}{2L}\right)^2\right]\,,
\label{e-modularity}
\end{equation}
where $N_M$ is the number of non-empty modules (smaller than or equal
to the number $N$ of nodes in the network), $L$ is the number of links
in the network, $l_{s}$ is the number of links between nodes in module
$s$, and $d_s$ is the sum of the degrees of the nodes in module
$s$. The objective of a module identification algorithm is to find the
partition $\mathcal{P^*}$ that yields the largest modularity
$M\equiv\mathcal{M(P^*)}$. Note that $N_M$ is only constrained to be
$N_M \le N$, but is otherwise selected by the optimization algorithm
so that $\mathcal{M}$ is maximum. The problem of identifying the
optimal partition is analogous to finding the ground state of a
disordered system with Hamiltonian
$\mathcal{H}=-L\mathcal{M}$.~\cite{guimera04}

Since the modularity landscape is in general very rugged, we use
simulated annealing to find a close to optimal partition of the
network into modules~\cite{guimera05a,guimera05,guimera04}. This
method is the most accurate to date~\cite{guimera05a,danon05}.


\subsection*{Role definition}

We determine the role of each node according to two properties
\cite{guimera05a,guimera05}: the relative within-module degree $z$ and
the participation coefficient $P$. The within-module degree $z$-score
measures how ``well-connected'' node $i$ is to other nodes in the
module compared to those other nodes, and is defined as
\begin{equation}
z_i = \frac{\kappa^i_{s_i} - \langle{\kappa}^j_{s_i} \rangle_{j \in
s_i}} {\sqrt{\langle ( \kappa^j_{s_i} )^2 \rangle_{j \in s_i} -
\langle{\kappa}^j_{s_i} \rangle_{j \in s_i}^2 }} \;,
\end{equation}
where $\kappa^i_s$ is the number of links of node $i$ to nodes in
module $s$, $s_i$ is the module to which node $i$ belongs, and the
averages $\langle \dots \rangle_{j \in s}$ are taken over all nodes in
module $s$.

The participation coefficient quantifies to what extent a node
connects to different modules We define the participation coefficient
$P_i$ of node $i$ as
\begin{equation}
P_i=1-\sum_{s=1}^{N_M}\left(\frac{\kappa^i_s}{k_i} \right)^2
\end{equation}
where $\kappa^i_s$ is the number of links of node $i$ to nodes in
module $s$, and $k_i=\sum_s \kappa^i_s$ is the total degree of node
$i$. The participation coefficient of a node is therefore close to one
if its links are uniformly distributed among all the modules and zero
if all its links are within its own module.

We classify as {\it non-hubs} those nodes that have low within-module
degree ($z<2.5$). Depending on the amount of connections they have to
other modules, non-hubs are further subdivided
into~\cite{guimera05a,guimera05}: (R1) {\it ultra-peripheral nodes},
that is, nodes with all their links within their own module ($P \le
0.05$); (R2) {\it peripheral nodes}, that is, nodes with most links
within their module ($0.05 < P \le 0.62$); (R3) {\it satellite
connectors}, that is, nodes with a high fraction of their links to
other modules ($0.62 < P \le 0.80$); and (R4) {\it kinless nodes},
that is, nodes with links homogeneously distributed among all modules
($P > 0.80$). We classify as {\it hubs} those nodes that have high
within-module degree ($z\ge 2.5$). Similar to non-hubs, hubs are
divided according to their participation coefficient into: (R5) {\it
provincial hubs}, that is, hubs with the vast majority of links within
their module ($P \le 0.30$); (R6) {\it connector hubs}, that is, hubs
with many links to most of the other modules ($0.30 < P \le 0.75$);
and (R7) {\it global hubs}, that is, hubs with links homogeneously
distributed among all modules ($P > 0.75$).


\subsection*{Network randomization and statistical ensembles}

We use two different ensembles of random
networks~\cite{maslov02,maslov04}. In the first ensemble, which we
denote by $\mathcal{D}$, we only preserve the degree sequence of the
original network; in the second ensemble, denoted $\mathcal{M}$, we
preserve both the degree sequence and the modular structure of the
network. Averages over the first and second ensembles are denoted
$\langle \dots \rangle_\mathcal{D}$ and $\langle \dots
\rangle_\mathcal{M}$, respectively.

To generate random networks in ensemble $\mathcal{D}$, we randomize
all the links in the network while preserving the degree of each
node. To uniformly sample all possible networks, we use the
Markov-chain Monte Carlo switching
algorithm~\cite{maslov02,itzkovitz04}. In this algorithm, one
repeatedly selects random pairs of links, for example $(i,j)$ and
$(l,m)$, and swaps one of the ends of each link, so that the links
become $(i,m)$ and $(l,j)$.

To generate random networks in ensemble $\mathcal{M}$, we restrict the
Markov-chain Monte Carlo switching algorithm~\cite{maslov04} to pairs
of links that connect nodes in the same pair of modules, that is, we
apply the Markov-chain Monte Carlo switching algorithm independently
to links whose ends are in modules 1 and 1, 1 and 2, and so forth for
all pairs of modules. This method guarantees that, with the same
partition as the original network, the modularity of the randomized
network is the same as that of the original network (since the number
of links between each pair of modules is unchanged) and that the role
of each node is also preserved.

To investigate whether global properties are representative of
module-specific properties, we focus on degree $k_i$, clustering
coefficient $C_i$, and normalized clustering coefficient $C_i /
\langle C_i \rangle_\mathcal{D}$. For each module $s$ in the network,
comprising $n_s$ nodes, we compute the average of each property in the
module (for example, $\langle k_i \rangle_{i \in s}$). Additionally,
we compute the distribution of such averages for random modules, which
we obtain by randomly selecting groups of $n_s$ nodes. If the
empirical module average falls outside of the 95\% probability of the
distribution for the random modules, we consider that the global
average is not representative of the module average. We finally
compute the fraction $r$ of modules that are not properly described by
the global average.

To study degree-degree correlations, we consider the average degree
$k_{\rm nn}^i$ of the nearest neighbors of each node $i$. We define
the normalized nearest neighbors' degree $d^i$ as the ratio of $k_{\rm
nn}^i$ and: %
(i) the average value of $k_{\rm nn}^j$ in the network
\begin{equation}
d^i_\mathcal{N} = \frac{N \, k_{\rm nn}^i}{\sum_j k_{\rm nn}^j}\; ,
\end{equation}
where $N$ is the number of nodes in the network;
(ii) the expected value of $k_{\rm nn}^i$ in the ensemble of networks
with fixed degree sequence
\begin{equation}
d^i_\mathcal{D} = \frac{k_{\rm nn}^i}{\langle k_{\rm nn}^i
\rangle_\mathcal{D}}\; ;
\end{equation}
and (iii) the expected value of $k_{\rm nn}^i$ in the ensemble of
networks with fixed degree sequence and modular structure
\begin{equation}
d^i_\mathcal{M} = \frac{k_{\rm nn}^i}{\langle k_{\rm nn}^i
\rangle_\mathcal{M}}\;.
\end{equation}
Note that, in spite of the similar notation, the meaning of
$d^i_\mathcal{N}$ is somewhat different from the other two because the
normalization involves an average over nodes, while in
$d^i_\mathcal{D}$ and $d^i_\mathcal{M}$ the normalization involves
averages over an ensemble of randomized networks.

To obtain the role-to-role connectivity profiles, we calculate the
$z$-score~\cite{maslov02,milo02,maslov04,milo04} of the number of
links between nodes with roles $i$ and $j$ as
\begin{equation}
  z_{ij} = \frac{r_{ij} - \langle r_{ij} \rangle_\mathcal{M}}
  {\sqrt{\langle r_{ij}^2 \rangle_\mathcal{M} - \langle r_{ij}
  \rangle_\mathcal{M}^2}} \;,
\end{equation}
where $r_{ij}$ is the number of links between nodes with roles $i$ and
$j$. To obtain better statistics and an estimation of the error in the
$z$-score, we carry out this process for several partitions of each
network.

To evaluate the similarity between two $z$-score profiles $\vec{a}$
and $\vec{b}$, we use the scalar product
\begin{equation}
  r_{ab} = \sum_{i, j\ge i} \frac{z_{ij}^a\, z_{ij}^b}{\sigma_{z^a} \,
  \sigma_{z^b}} \; ,
\end{equation}
where $\sigma_{z^a}$ is the standard deviation of the elements in
$\vec{a}$.


%% file: table-datasets.tex
\begin{table}
\begin{center}
\begin{tabular}{llccccc}
\hline
Network type & Network & Nodes & Links & $N_M$ & $M$ & $\langle M \rangle_\mathcal{D}$ \\
\hline
\multirow{6}{*}{Metabolism Archaea} & {\it A. fulgidus}  & 303 & 366 & 16 & 0.813 & 0.746 (0.005) \cr
& {\it A. pernix}  & 300 & 387 & 14 & 0.797 & 0.711 (0.006) \cr
& {\it M. jannaschii}  & 223 & 277 & 14 & 0.813 & 0.720 (0.003) \cr
& {\it P. aerophilum}  & 335 & 421 & 15 & 0.811 & 0.731 (0.004) \cr
& {\it P. furiosus}  & 302 & 384 & 16 & 0.813 & 0.720 (0.007) \cr
& {\it S. solfataricus}  & 367 & 455 & 17 & 0.813 & 0.736 (0.006) \cr
\hline
\multirow{6}{*}{Metabolism Bacteria} & {\it B. subtilis}  & 649 & 863 & 20 & 0.815 & 0.724 (0.003) \cr
& {\it E. coli}  & 739 & 1009 & 17 & 0.810 & 0.711 (0.003) \cr
& {\it F. nucleatum}  & 378 & 473 & 16 & 0.816 & 0.734 (0.004) \cr
& {\it H. pylory}  & 360 & 438 & 15 & 0.837 & 0.746 (0.006) \cr
& {\it M. leprae}  & 451 & 578 & 16 & 0.814 & 0.732 (0.005) \cr
& {\it T. elongatus}  & 448 & 546 & 17 & 0.830 & 0.755 (0.006) \cr
\hline
\multirow{6}{*}{Metabolism Eukaryotes} & {\it A. thaliana}  & 607 & 792 & 18 & 0.825 & 0.728 (0.003) \cr
& {\it C. elegans}  & 431 & 569 & 17 & 0.818 & 0.714 (0.004) \cr
& {\it H. sapiens}  & 792 & 1056 & 23 & 0.842 & 0.727 (0.003) \cr
& {\it P. falciparum}  & 280 & 363 & 12 & 0.815 & 0.708 (0.006) \cr
& {\it S. cerevisiae}  & 570 & 776 & 17 & 0.814 & 0.708 (0.003) \cr
& {\it S. pombe}  & 503 & 664 & 18 & 0.827 & 0.721 (0.003) \cr
\hline
\multirow{3}{*}{Air transportation} & Global & 3618 & 14142 & 25 & 0.706 & 0.3111 (0.0009) \cr
& Asia \& Middle East & 706 & 2572 & 10 & 0.642 & 0.325 (0.002) \cr
& North America & 940 & 3446 & 12 & 0.522 & 0.3111 (0.0005) \cr
\hline
\multirow{2}{*}{Interactome} & {\it S. cerevisiae} & 1458 & 1948 & 25 & 0.820 & 0.707 (0.002) \cr
& {\it C. elegans} & 2889 & 5188 & 28 & 0.688 & 0.561 (0.002) \cr
\hline
\multirow{3}{*}{Internet} & 1998 & 3216 & 5705 & 17 & 0.625 & 0.5365 (0.0011) \cr
& 1999 & 4513 & 8374 & 18 & 0.620 & 0.5227 (0.0007) \cr
& 2000 & 6474 & 12572 & 22 & 0.631 & 0.5042 (0.0008) \cr
\hline
\end{tabular}
\caption{
Properties and modularity of the studied networks. We show the number
of nodes and links in the network, the modularity $M$ of the best
partition obtained using simulated annealing, and the average
modularity $\langle M \rangle_\mathcal{D}$ (and standard deviation) of
the randomizations of the network, obtained using the Markov-chain
switching algorithm to preserve the degree of each node (see
Methods). Note that all networks are significantly modular, that is,
their modularity is larger than the modularity of their corresponding
randomizations.
}
\label{t-datasets}
\end{center}
\end{table}

%% file: table-module-globprop-likelyhood.tex
\begin{table}[h!]
\begin{center}
\begin{tabular}{llccc}
\hline
Network type & Network & $r_{\langle k_i \rangle_i}$ & $r_{\langle C_i \rangle_i}$ & $r_{\langle C_i / \langle C_i \rangle_\mathcal{D} \rangle_i}$ \cr
\hline
\multirow{6}{*}{Metabolism Archaea} & {\it A. fulgidus} & 0.02 (0.03) & 0.125 (0.0) & 0.10 (0.03) \cr
& {\it A. pernix} & 0.0 (0.0) & 0.17 (0.04) & 0.18 (0.04) \cr
& {\it M. jannaschii} & 0.0 (0.0) & 0.27 (0.03) & 0.27 (0.02) \cr
& {\it P. aerophilum} & 0.03 (0.03) & 0.22 (0.06) & 0.16 (0.05) \cr
& {\it P. furiosus} & 0.02 (0.03) & 0.27 (0.04) & 0.24 (0.06) \cr
& {\it S. solfataricus} & 0.02 (0.03) & 0.15 (0.04) & 0.11 (0.04) \cr
\hline
\multirow{6}{*}{Metabolism Bacteria} & {\it B. subtilis} & 0.02 (0.02) & 0.22 (0.06) & 0.19 (0.04) \cr
& {\it E. coli} & 0.02 (0.04) & 0.27 (0.06) & 0.29 (0.04) \cr
& {\it F. nucleatum} & 0.0 (0.0) & 0.06 (0.02) & 0.06 (0.03) \cr
& {\it H. pylori} & 0.08 (0.05) & 0.28 (0.04) & 0.26 (0.03) \cr
& {\it M. leprae} & 0.0 (0.0) & 0.28 (0.05) & 0.27 (0.04) \cr
& {\it T. elongatus} & 0.01 (0.02) & 0.11 (0.03) & 0.12 (0.04) \cr
\hline
\multirow{6}{*}{Metabolism Eukaryotes} & {\it A. thaliana} & 0.04 (0.03) & 0.29 (0.06) & 0.29 (0.07) \cr
& {\it C. elegans} & 0.064 (0.004) & 0.31 (0.03) & 0.30 (0.03) \cr
& {\it H. sapiens} & 0.08 (0.03) & 0.45 (0.04) & 0.41 (0.05) \cr
& {\it P. falciparum} & 0.084 (0.002) & 0.23 (0.03) & 0.24 (0.02) \cr
& {\it S. cerevisiae} & 0.09 (0.04) & 0.24 (0.05) & 0.23 (0.05) \cr
& {\it S. pombe} & 0.059 (0.003) & 0.37 (0.06) & 0.36 (0.06) \cr
\hline
\multirow{3}{*}{Air transportation} & Global & 0.41 (0.05) & 0.531 (0.010) & 0.43 (0.02) \cr
& Asia \& Middle East & 0.40 (0.10) & 0.26 (0.04) & 0.21 (0.05) \cr
& North America & 0.37 (0.03) & 0.40 (0.04) & 0.47 (0.05) \cr
\hline
\multirow{2}{*}{Interactome} & {\it S. cerevisiae} & 0.0 (0.0) & 0.25 (0.09) & 0.67 (0.04) \cr
& {\it C. elegans} & 0.042 (0.014) & 0.47 (0.06) & 0.33 (0.04) \cr
\hline
\multirow{3}{*}{Internet} & 1998 & 0.064 (0.005) & 0.77 (0.05) & 0.77 (0.06) \cr
& 1999 & 0.0 (0.0) & 0.85 (0.03) & 0.83 (0.05) \cr
& 2000 & 0.0 (0.0) & 0.77 (0.04) & 0.76 (0.07) \cr
\hline
\end{tabular}
\caption{
Global versus module-specific average properties.
For each network, we show the fraction $r$ of modules (and standard
deviation) whose average degree $\langle k_i \rangle_i$, clustering
coefficient $\langle C_i \rangle_i$, and normalized clustering
coefficient $\langle C_i / \langle C_i \rangle_\mathcal{D} \rangle_i$
significantly differ (at a 95\% confidence) from the global network
average (Methods). Fractions $r>0.05$ indicate that a given global
property does not correctly describe individual modules. Global degree
is not representative of individual-module degrees for air
transportation networks. Most importantly, the global clustering
coefficient is not representative of individual-module clustering
coefficients for any network (except, maybe, the metabolic network of
{\it F. nucleatum}).
}
\label{t-globprop}
\end{center}
\end{table}